\documentclass[prl,amsmath,amssymb,reprint,superscriptaddress]{revtex4-1}
\usepackage{bm}
\usepackage{graphicx}

\begin{document}

%\definecolor{r}{rgb}{0.9,0,0}

\newcommand{\ie}{\emph{i.e.}}
\newcommand{\eg}{\emph{e.g.}}
\newcommand{\dtwomin}{D^2_\mathrm{min}}
\newcommand{\etal}{\emph{et al.}}

\title{Global Memory from Local Hysteresis in an Amorphous Solid}

\author{Nathan~C.~Keim}
\altaffiliation{Present address: Department of Physics, Pennsylvania State University, University Park, Pennsylvania 16802, USA}
\email{keim@psu.edu}
\affiliation{Department of Physics, California Polytechnic State University, San Luis Obispo CA 93407, USA}

\author{Jacob Hass}
\affiliation{Department of Physics, California Polytechnic State University, San Luis Obispo CA 93407, USA}

\author{Brian Kroger}
\affiliation{Department of Physics, California Polytechnic State University, San Luis Obispo CA 93407, USA}

\author{Devin Wieker}
\affiliation{Department of Physics, California Polytechnic State University, San Luis Obispo CA 93407, USA}

\date{\today}

\begin{abstract}

A disordered material that cannot relax to equilibrium, such as an amorphous or glassy solid, responds to deformation in a way that depends on its past. In experiments we train a 2D athermal amorphous solid with oscillatory shear, and show that a suitable readout protocol reveals the shearing amplitude. When shearing alternates between two amplitudes, signatures of both values are retained only if the smaller one is applied last. We show that these behaviors arise because individual clusters of rearrangements are hysteretic and dissipative, and because different clusters respond differently to shear. These roles for hysteresis and disorder are reminiscent of the return-point memory seen in ferromagnets and many other systems. Accordingly, we show how a simple model of a ferromagnet can reproduce key results of our experiments and of previous simulations. Unlike ferromagnets, amorphous solids' disorder is unquenched; they require ``training'' to develop this behavior.
\end{abstract}

\maketitle

We are familiar with our own memory and forgetfulness, and digital memories are woven into our lives. But throughout our environment, matter is being driven without relaxing to equilibrium, potentially forming memories of its own: specific information about past conditions that can be recalled later. As a simple example, rubber ``remembers'' the extrema of all deformations since it was cured~\citep{Schmoller:2013ir}; the material stiffens as it is driven beyond those limits, allowing the memory to be read. Further afield, dilute non-Brownian suspensions that are sheared cyclically~\cite{Keim:2011dv,Paulsen:2014hm} and charge density-wave conductors given electrical pulses~\cite{Coppersmith:1997cj,Povinelli:1999jp} share distinctive rules for remembering multiple input values. Studying memory can thus reveal unexpected connections between systems and prompt new examinations of their physics~\cite{Keim:2019aa}.

\begin{figure}
\includegraphics[width=3.3in]{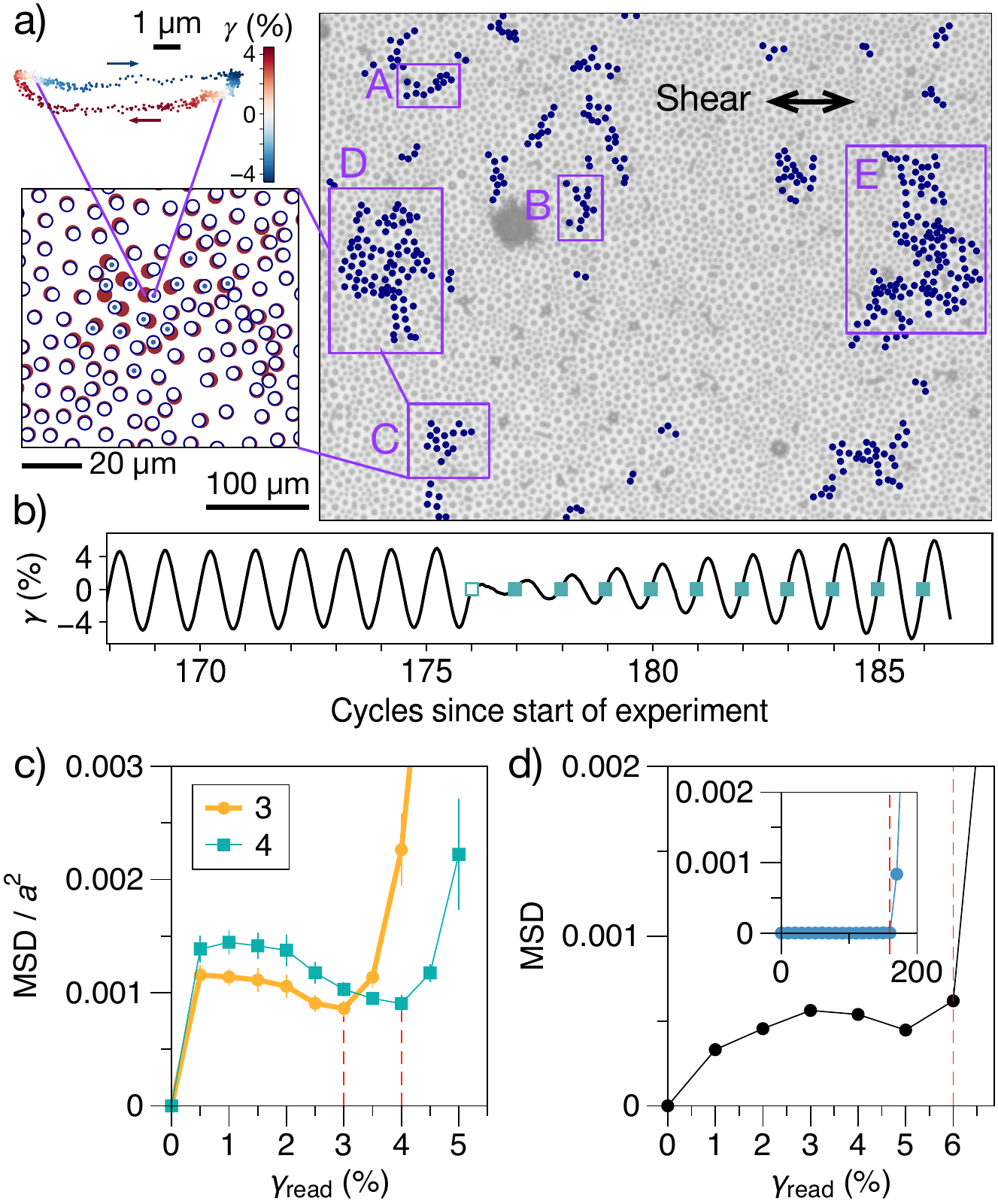}
\caption{
\label{fig:readexp}
Training and readout in experiments.
\textbf{(a)} Upper left: Positions of a single particle at 600 times in a cycle of shear. Colors correspond to global shear strain $\gamma$. Loop structure indicates hysteresis. Lower left: A cluster of rearrangements. Red filled/blue open circles mark particle positions when $\gamma = 0$, for increasing/decreasing $\gamma$. Dots mark particles identified by our analysis as rearranging. Right: Portion of the material, with rearranging particles marked. Labeled boxes identify several clusters of rearrangements, further analyzed in Fig.~\ref{fig:activity}. 
\textbf{(b)} Shear strain. Training is with constant amplitude $\gamma_1 = 4\%$. The leftmost, open symbol marks the trained state. During readout, the system's state at each time marked with a symbol is compared with the trained state.
\textbf{(c)} Normalized mean-squared displacement of all particles during readout, as a function of the strain amplitude just applied. Legend indicates training amplitude (in \%). Error bars represent standard deviation of mean for several trials~\cite{suppmat}.
\textbf{(d)} Analogous readout of 6\% training in simulations of an amorphous solid, adapted from Adhikari and Sastry~\cite{Adhikari:2018il}. 
\textit{Inset:} Analogous readout in dilute suspension experiments by Paulsen et al.~\cite{Paulsen:2014hm}.
}
\end{figure}

Recently, a new memory behavior was discovered in amorphous solids~\citep{Fiocco:2014bz}. This vast class of materials features atoms or particles packed with a minimum of the regular placement found in crystals. Amorphous solids made of molecules, bubbles, macroscopic grains, or colloidal particles (Fig.~\ref{fig:readexp}a) deform in remarkably similar ways: applied stress tends to cause localized clusters of particles (``soft spots'') to rearrange, marking transitions among a vast set of metastable states~\cite{Falk:2011co,Cubuk:2017bq,Manning:2011ha}. Yet under oscillatory shear, after many cycles these rearrangements can become periodic; particles' trajectories become loops~\citep{Regev:2013es,Keim:2013je,Nagamanasa:2014jx,Keim:2014hu,Fiocco:2014bz,Regev:2015hs}. Molecular dynamics simulations of glasses~\citep{Fiocco:2014bz,Fiocco:2015kr,Adhikari:2018il} and experiments on bubble rafts~\cite{Mukherji:2019hp} showed that after a strain amplitude $\gamma_1$ has been applied repeatedly to reach a ``trained'' steady state, the material retains an imprint of its training: a readout protocol can reveal $\gamma_1$. This protocol is illustrated in Fig.~\ref{fig:readexp}b: cycles of increasing amplitude $\gamma_\text{read}$ are applied, beginning with an amplitude below the training value~\cite{Keim:2011dv,Paulsen:2014hm, Mukherji:2019hp}. After each cycle, one measures the mean squared displacement (MSD) of the particles, relative to the trained state. A local minimum in the MSD as a function of $\gamma_\text{read}$ shows evidence of the training amplitude, as in Figs.~\ref{fig:readexp}c (our experiments) and d (simulations of Adhikari and Sastry~\cite{Adhikari:2018il}). If the same procedure is performed with a dilute non-Brownian suspension, the data reveal the training amplitude in a different way: the MSD is negligible until $\gamma_\text{read}$ exceeds the training amplitude (inset of Fig.~\ref{fig:readexp}d) \cite{Paulsen:2014hm}.

These findings represent new possibilities for describing and exploiting these materials' complex history-dependence, but they also prompt new questions: What is the mechanism for memory formation and readout? What can memory reveal about the physics of amorphous solids more broadly? How should one place this behavior among examples of memory in other systems, and what explains the contrast with a more dilute system of particles?

% {{{ Figure
%}}}

In this paper, we describe experiments with the two-dimensional amorphous solid in Fig.~\ref{fig:readexp}a, showing the readout of stored memories, consistent with other systems \citep{Fiocco:2014bz,Fiocco:2015kr,Adhikari:2018il,Mukherji:2019hp}. We propose that these memory results are approximately consistent with a behavior called return-point memory (RPM) that is exhibited by many hysteretic systems~\cite{barker83,Sethna:1993ts,Keim:2019aa}. We use a simple model with RPM to illustrate the basic mechanism. Finally, we return to the experimental system to identify this mechanism at work in the hysteresis of rearranging particles. Our findings help to explain this memory behavior and why it is different from that of dilute suspensions, and suggests that the material must be ``trained'' to behave this way.

\subsection{Methods}

Our experiment consists of polystyrene sulfate latex particles (Invitrogen), with diameters 3.7~$\mu$m (Lot 1839598) and 5.4~$\mu$m (Lot 1818113) in roughly equal numbers (Fig.~\ref{fig:readexp}a), adsorbed at the interface between decane (``99\%+,'' ACROS Organics) and deionized water in a 60~mm-diameter glass dish~\citep{Keim:2014hu}. The particle suspension is handled using pipette tips and Eppendorf tubes that are free of surface treatments (Axygen ``Maxymum Recovery'') and it includes 50\% ethanol as a spreading agent. 

These particles exhibit long-range electrostatic repulsion~\cite{Masschaele:2010da}, and so at the concentrations used here (area fraction $0.36 \pm 0.04$ \cite{suppmat}) each particle is mechanically over-constrained by its neighbors but does not touch them---forming a soft, frictionless jammed 2D solid with a typical spacing $a=8.2$~$\mu$m between particle centers, as measured at the first peak of the pair correlation function $g(r)$~\cite{trackpyv04}. 

\begin{figure}
    \begin{center}
        \includegraphics[width=2in]{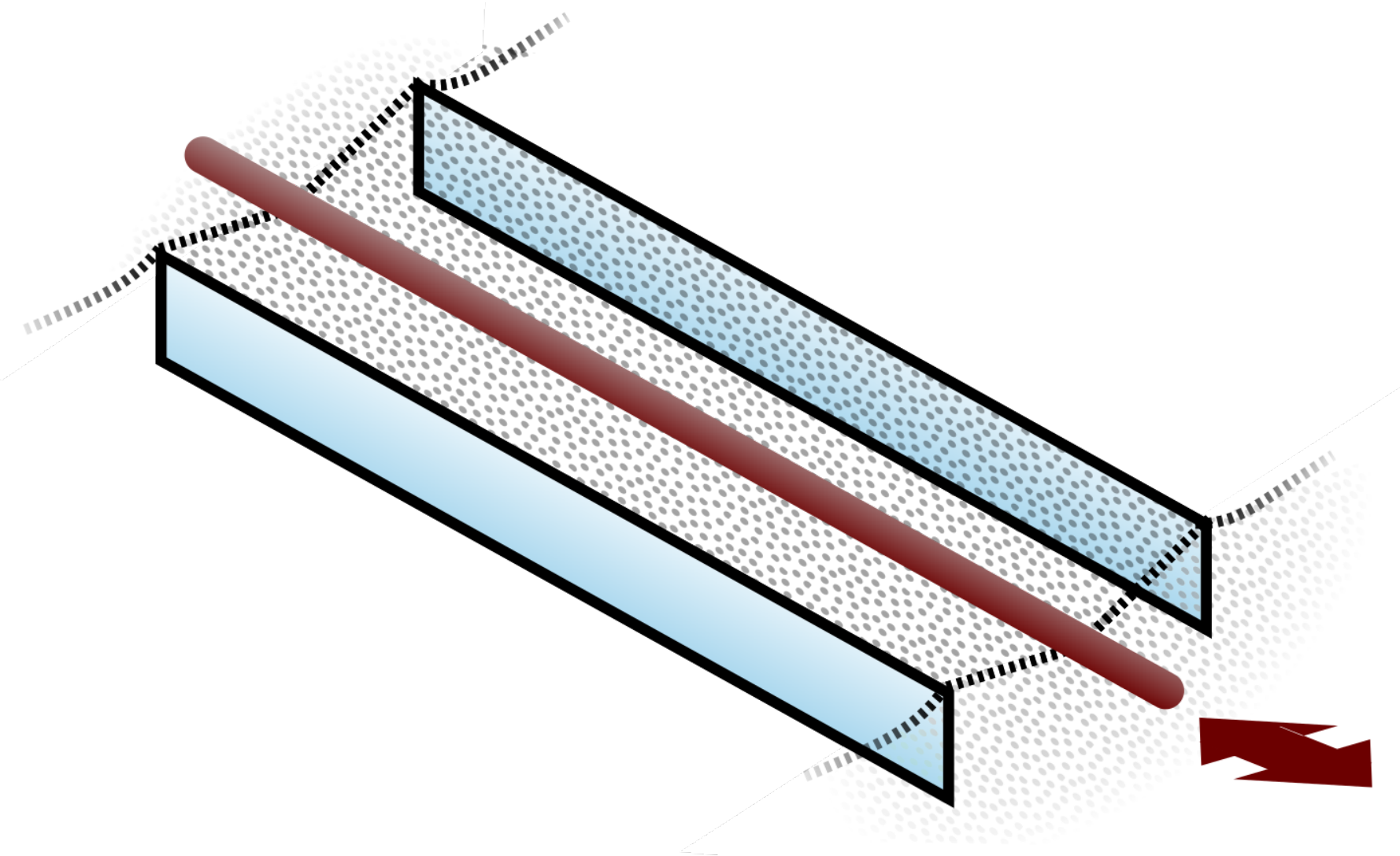}
    \end{center}
    \caption{Schematic of the interfacial stress rheometer apparatus. A magnetized needle is adsorbed at the interface, along with the particle monolayer, in a channel formed by two 18~mm square glass coverslips that are held 3.2~mm apart by a nylon clamp (not shown). Dashed lines indicate that the meniscus rises slightly at the walls. Arrow indicates the direction of magnetically-driven needle motion, parallel to the channel.
        \label{fig:apparatus}
}
\end{figure}

We use an interfacial shear rheometer~\citep{Brooks:1999ky,Reynaert:2008dm,Keim:2014hu} to shear the material. As shown in Fig.~\ref{fig:apparatus}, a steel needle is adsorbed at the interface and is positioned by a magnetic field between two glass walls; a computer-controlled perturbative field then drives the needle sinusoidally. Shear is nearly uniform, due to the no-slip boundary conditions at the needle and walls, and the high ratio of interfacial shear stress to bulk liquid (oil and water) shear stress---the Boussinesq number $\text{Bq} \sim 100$~\cite{Reynaert:2008dm}. The needle is 230~$\mu$m in diameter and 32~mm long; it protrudes from the ends of the channel to keep this yield-stress material from forming solid ``plugs'' there. This means that the working sample is approximately 18~mm long and 1.5~mm wide on each side of the needle. 

Synchronously with shearing, we image and track $\sim$40,000 particles in a $1.9 \times 1.4$~mm area~\citep{Crocker:1996wp,trackpyv04}. We use a long-distance microscope (Infinity K2/SC) and 4-megapixel machine-vision camera (Mikrotron 4CXP) at a magnification of 0.82~$\mu$m/pixel and a frame rate of 30 frames/s. To find particle locations, we first apply a binary threshold to each image and estimate the locations of large particles. This allows us to compute the precise centroids of particles in two passes, optimized separately for large and small particles. High-throughput tracking is performed with the open-source ``trackpy'' software~\cite{trackpyv04,Crocker:1996wp} using the channel-flow prediction and adaptive search features, with the help of an image-registration algorithm to compensate for occasional motions of the microscope due to external vibrations. To reduce the effect of spurious rearrangements caused by particle-tracking errors, we discard any particle that is not tracked continuously over an entire set of samples, \emph{e.g.}\ the entire readout process.

Analysis involves measuring the differences between particle positions at two different times. For each particle, we subtract the average motion of a region of nearby material (radius $R_\text{disp} = 8.5a$), to avoid spurious signals due to small motions of the camera or variation of the needle position, yielding $\Delta \vec r_\text{local}$~\cite{Keim:2014hu,philatracksv02}. Choosing $R_\text{disp} = 4.5a$ or $16.5a$ does not change our qualitative results~\cite{suppmat}.

\subsection{Training and Readout}

\begin{figure}
\includegraphics[width=3.43in]{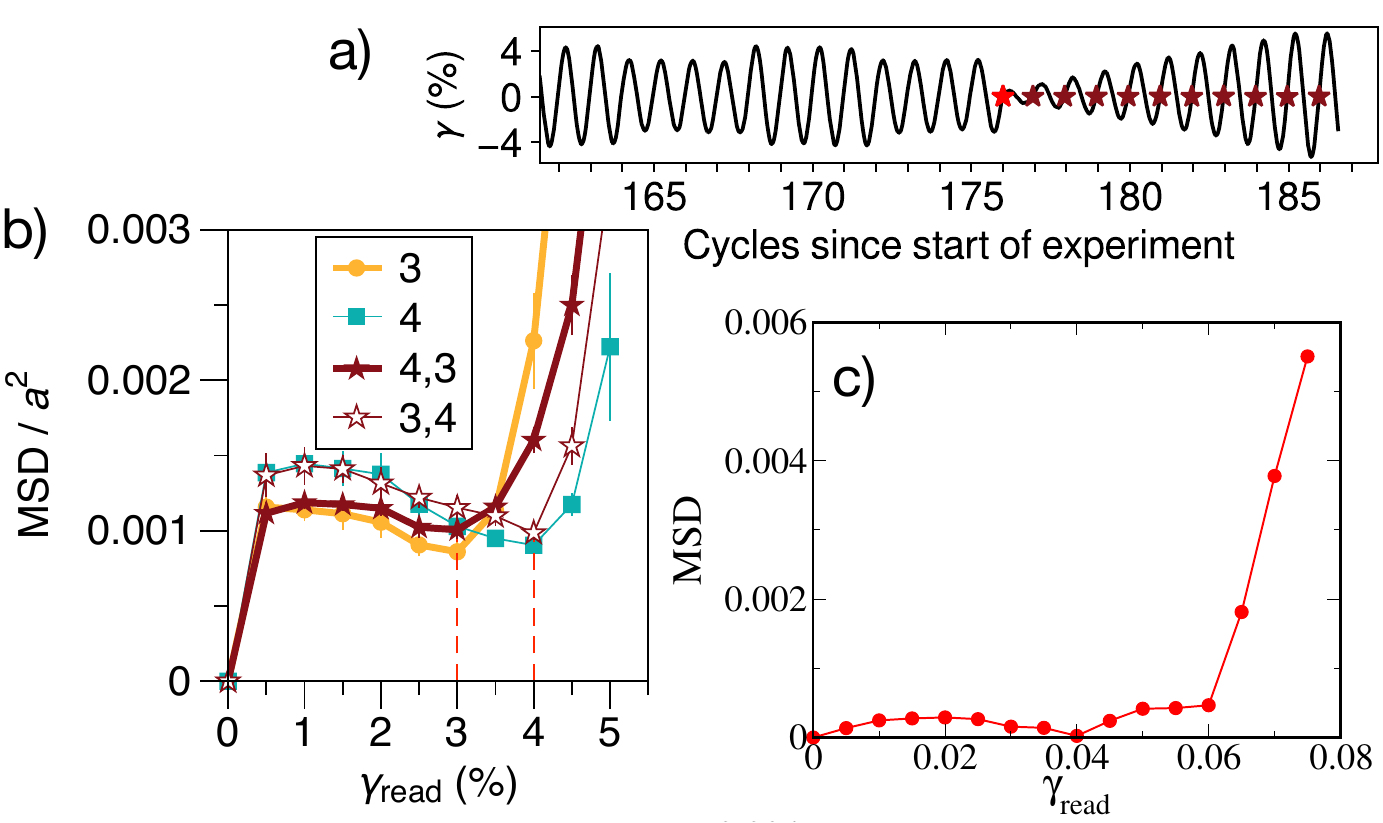}
\caption{
\label{fig:read2}
Two memories.
\textbf{(a)} Experimental protocol as in Fig.~\ref{fig:readexp}b, but with $\gamma_2 = 4\%, \gamma_1 = 3\%$.
\textbf{(b)} Experimental readout for all particles as in Fig.~\ref{fig:readexp}c. ``$4, 3$'' is the protocol in panel (a) and shows evidence of both memories; ``$3,4$'' indicates $\gamma_2 = 4\%$ was applied last before readout, erasing the memory at 3\%.
\textbf{(c)} Simulations by Adhikari and Sastry~\cite{Adhikari:2018il}, after training with 6\% and 4\%.
}
\end{figure}

All of the experiments reported here follow the protocol: (1) a ``reset'' phase where we apply 6 cycles with strain amplitude $\sim 70\%$ at 0.1 Hz; (2) a ``training'' phase where we apply oscillatory shear at 0.05 Hz with a repeating pattern of strain amplitudes for 176 cycles, recording video for the last 24; (3) a ``readout'' phase. Figure~\ref{fig:readexp}a shows strain \emph{vs.}\ time at the end of one experiment.  Training involves the pattern of amplitudes $\gamma_2$, $\gamma_2$, $\gamma_2$, $\gamma_2$, $\gamma_1$, $\gamma_1$, $\gamma_1$, $\gamma_1$ (176 cycles = 22 repetitions). We use $\gamma_1 = \gamma_2$ (Fig.~\ref{fig:readexp}), as well as $\gamma_1 < \gamma_2$ and $\gamma_1 > \gamma_2$ (Fig.~\ref{fig:read2}). $\gamma_1$ is always applied \emph{last} before readout. Amplitudes are repeated within the pattern to reduce the possibility that the material would ``learn'' a 2-cycle trajectory, in which the amplitude of one cycle always predicts the amplitude of the next. The duration of training is much longer than the $\sim$15 cycles typically required to reach an apparent steady state~\cite{Keim:2013je,Keim:2014hu}, so that by the end of training virtually all particles return to the same positions after a complete 8-cycle pattern, despite many rearrangements (Fig.~\ref{fig:readexp}a). The median normalized MSD after 8 cycles in the steady state is $0.0010$, which is a scale for the noise floor in measurements like Fig.~\ref{fig:readexp}c. 

% {{{ Figure
%}}}

Consistent with other studies~\cite{Fiocco:2014bz,Fiocco:2015kr,Adhikari:2018il,Mukherji:2019hp}, we see evidence for both single and multiple memories. When we train with both 3\% and 4\% strain, applying $\gamma_1 = 3\%$ last before readout (Fig.~\ref{fig:read2}a, and ``4, 3'' curve in Fig.~\ref{fig:read2}b), we observe a memory at $\gamma_\text{read} = 3\%$, but we also see evidence for a memory above 3\%: MSD in that region is distinct from the ``3'' curve. The result is very different when we exchange $\gamma_1, \gamma_2$ and apply the larger amplitude last (``3, 4''): the signature of the smaller value is gone, which differs from the expected behavior of a dilute suspension~\cite{Keim:2011dv,Paulsen:2019ki,Keim:2019aa}. These results bear a resemblance to return-point memory (RPM). In the present context, RPM means that a cycle with amplitude $\gamma_1$ restores the system to the state it had after the previous cycle with $\gamma_1$ (i.e., minimizes MSD), so long as strain did not exceed $\gamma_1$ in the interim (the difference between ``4, 3'' and ``3, 4'' training)~\cite{barker83,Sethna:1993ts,Paulsen:2019ki,Keim:2019aa}. 
In the rest of this paper, we explore the possibility that RPM could at least partially explain memory in amorphous solids.

\subsection{Model and Mechanism}

% {{{ Figure
\begin{figure}
\includegraphics[width=3.43in]{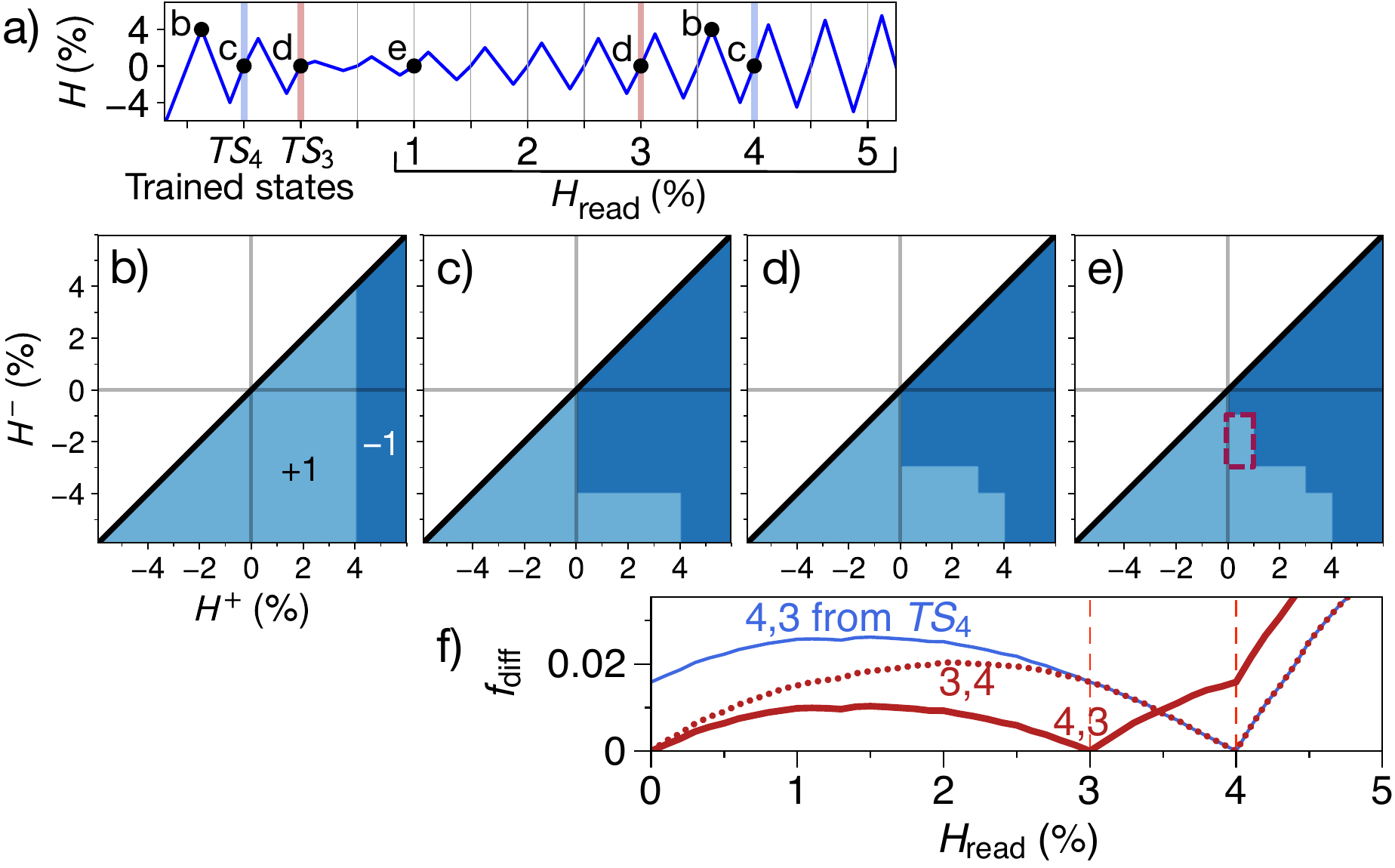}
\caption{
\label{fig:preisach}
The Preisach model illustrates how specific memories can emerge from hysteresis.
\textbf{(a)} Applied field $H$ is varied to store and read memories of 4\% (trained state TS$_4$) and 3\% (TS$_3$). Labels correspond to panels \textbf{(b)--(e)}, which show the model's many hysterons, plotted according to  $H^+$, $H^-$. Increasing $H$ converts hysterons to the $+1$ state, growing the lighter-shaded region rightward; decreasing $H$ grows the $-1$ region downward. For example, to go from (b) to (c) we decrease $H$ from 4\% to $-4\%$, then increase to 0.
In (e), a cycle with $\gamma_\text{read} = 1\%$ flipped hysterons in the outlined region to $+1$, but did not flip them back,causing a difference with the trained state (d).  
\textbf{(f)} A simulated Preisach model is read out by measuring the fraction of hysterons $f_\text{diff}$ that differ from a trained state. Measuring relative to TS$_3$ yields both memories (the ``4,3'' curve). Measuring relative to TS$_4$ (``4,3 from TS$_4$''), or applying the 4\% amplitude last before readout (``3,4''), shows a 4\% memory only.
}
\end{figure}
%}}}

We illustrate the mechanics of RPM with the Preisach model, originally used to study hysteresis in ferromagnets~\cite{barker83, Preisach:1935jm}. It considers many hysteretic subsystems, or ``hysterons," that are coupled to an external field $H$. The $i$th hysteron will ``flip" from its --1 state to +1 when $H$ is increased past $H_i^+$; it will flip back to --1 when $H$ is decreased past $H_i^-$. $H_i^+$ and $H_i^-$ are distributed uniformly from $-0.1$ to $0.1$ to represent disorder, and $H_i^+ > H_i^-$ to represent dissipative dynamics. We apply the training and readout protocol in Fig.~\ref{fig:preisach}a, and monitor hysterons' states in Fig.~\ref{fig:preisach}(b--e). The figure shows that our earlier definition of RPM is recursive: when $H_\text{read} = 4\%$, we recover the same state (b) as when amplitude 4\% was last applied, regardless of the intervening storage and recovery of a 3\% memory. In effect, there are \emph{two} trained states, which we denote TS$_3$ and TS$_4$. 

Figure~\ref{fig:preisach}e highlights hysterons that because of their $H^+$, are placed in the +1 state by applying $H \ge 1\%$; but because of their $H^-$, require $H \le -3\%$ to be fully reversed. During a readout cycle with amplitude $H_\text{read} \ge 3\%$, these hysterons would each flip to $+1$ and back to $-1$, but with $H_\text{read} = 1\%$, they are stuck in their +1 states. Extended to all hysterons, this basic mechanism of RPM means that reducing the driving amplitude leaves the entire system in a different state, but it also means that previous states TS$_3$ and TS$_4$ can be restored by increasing the amplitude to previous values.

Figure~\ref{fig:preisach}f shows that the readout protocol in Figs.~\ref{fig:readexp}c and \ref{fig:read2}a can also read RPM in a simulated Preisach model with 25,000 hysterons, with non-monotonic curves as in the experimental results. Instead of MSD, at the end of each cycle we measure the fraction of hysterons $f_\text{diff}$ that do not match a trained state. In the ``4, 3'' curve, the change in slope as $H_\text{read}$ passes 4\% comes from the many hysterons with $H^+ > 4\%$ or $H^- < -4\%$ that were heretofore inactive. Figure~\ref{fig:preisach}f also verifies the recursive nature of RPM with respect to TS$_3$ and TS$_4$.
These curves roughly match our experiments, and are strikingly similar to results from molecular dynamics simulations (Fig.~\ref{fig:read2}b)~\cite{Fiocco:2014bz, Fiocco:2015kr, Adhikari:2018il}, despite key differences in these systems' physics that we discuss below. (In recent bubble raft experiments~\cite{Mukherji:2019hp} the larger memory was instead observed as a second minimum, but the annular geometry in that work makes direct comparisons difficult.)

% {{{ Figure
\begin{figure}
\includegraphics[width=3.3in]{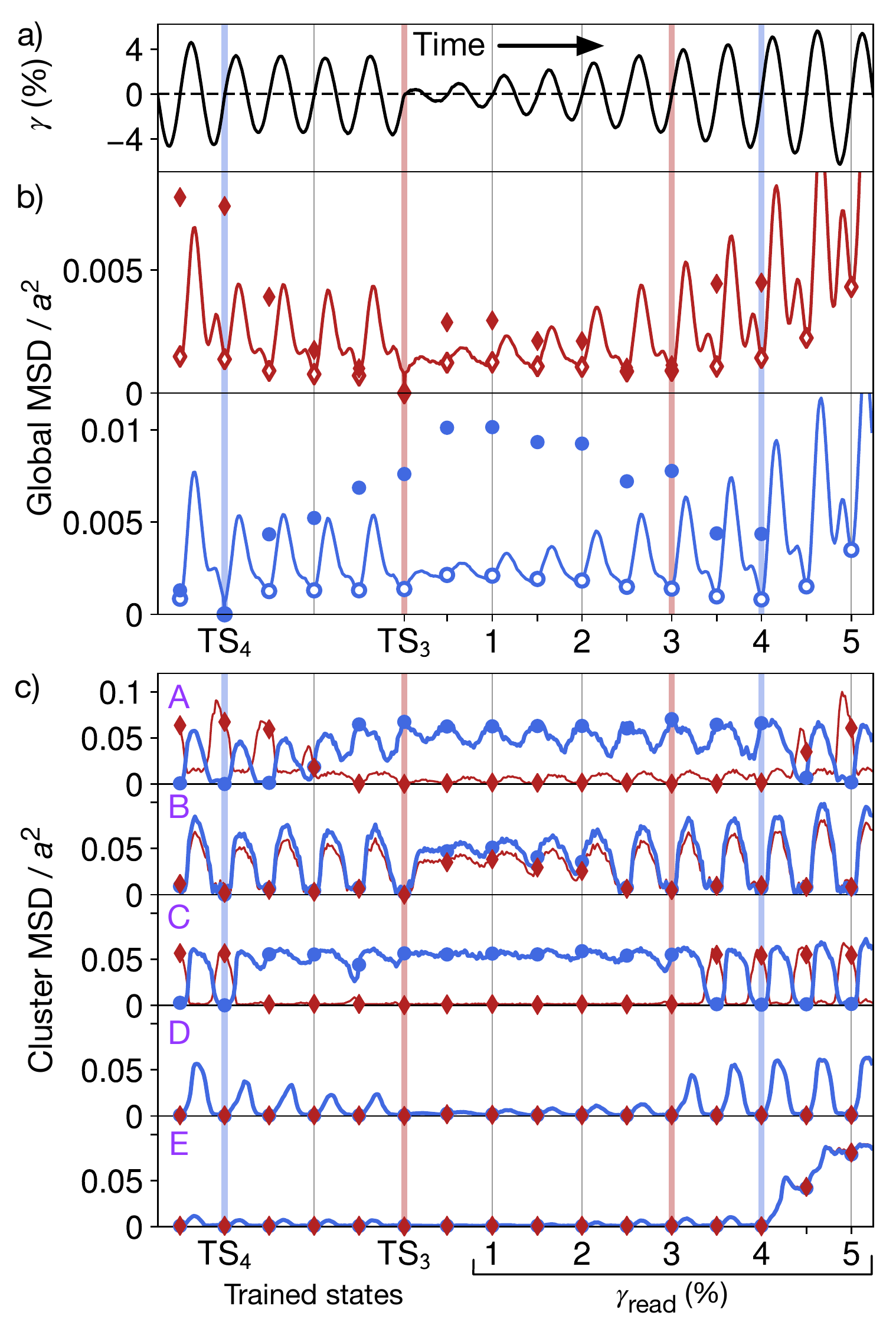}
\caption{
\label{fig:activity}
Training and readout at the level of individual rearrangements of particles, from one of the ``4, 3'' movies averaged in Fig.~\ref{fig:read2}b.
\textbf{(a)} Strain at end of training, and readout, as in Fig.~\ref{fig:read2}a. Pairs of thick blue and red vertical lines indicate trained states TS$_4$ and TS$_3$ that follow 4\% and 3\% cycles, respectively, and the corresponding times during readout (see axis at bottom of figure). 
\textbf{(b)} MSD of all particles in one movie, measured from 3\% (upper) and 4\% (lower) trained states. An open diamond or circle marks the value at the end of each cycle, as plotted in Fig.~\ref{fig:read2}. Closed symbols show values computed from only the 5 labeled clusters of rearranging particles in Fig.~\ref{fig:readexp}a; values for $\gamma_\text{read} > 4$ are too large to be plotted here. Both TS$_4$ and TS$_3$ are recovered approximately during readout.
\textbf{(c)} MSD of each labeled cluster in Fig.~\ref{fig:readexp}a, calculated relative to TS$_4$ (thick blue curve) and TS$_3$ (thin red curve). A blue circle and red diamond mark the end of each cycle. Each group's state at the end of a cycle is hysteretic, and depends on the strain amplitude in a different way. Collectively they give rise to the memory readouts in panel (b).
}
\end{figure}
%}}}

We now consider whether the mechanism illustrated by the Preisach model---hysteretic subsystems that can get stuck in one state when the driving amplitude is decreased---is relevant for our amorphous solid. We examine a movie with ``4, 3'' training in Fig.~\ref{fig:activity}a. To identify candidate subsystems of rearranging particles, we focus on the region in Fig.~\ref{fig:readexp}a (13\% of the recorded area), and compare each particle's position at the beginning of readout with its position at all other times that $\gamma = 0$ (twice per cycle), during the interval in Fig.~\ref{fig:activity}a. We mark a particle as rearranging (Fig.~\ref{fig:readexp}a) if its $\|\Delta \vec r_\text{local} / a\|^2 \ge 0.025$ in any sample~\cite{suppmat}.  To identify discrete subsystems, we use neighbor relationships (separation $< 1.5a$) to group these particles into contiguous rearranging clusters. Five clusters of interest are labeled A--E in Fig.~\ref{fig:readexp}a.

Figure~\ref{fig:activity}b shows global MSD of all particles (as in Fig.\ \ref{fig:read2}b) during training and readout, now computed 30 times per cycle, relative to two trained states: after the last application of $\gamma_1 = 3\%$ (TS$_3$, upper plot), and after the last application of $\gamma_2 = 4\%$ (TS$_4$, lower plot). However, we now also plot the MSD of the 5 labeled clusters only (closed symbols). This small fraction of the material is enough to qualitatively reproduce the global behavior. In Fig.~\ref{fig:activity}c we plot the MSD for each labeled cluster separately, relative to TS$_3$ (thin curves, diamonds) and TS$_4$ (thick curves, circles).

Figure~\ref{fig:activity}c shows that the global memory arises from local hysteresis and disorder. Because of disorder, each cluster rearranges at a different value of global strain, and so each cluster plays a different role in storing and reading memories. For instance, cluster ``B'' contributes strongly to the memory of TS$_3$: when driving amplitude is reduced at the start of readout, ``B''  gets stuck in a rearranged state relative to TS$_3$, and does not switch fully back to its original state until $\gamma_\text{read} \gtrsim 2.5\%$. In this way, ``B'' plays the same role that the highlighted hysterons of Fig.~\ref{fig:preisach}e did in the Preisach model.

Likewise, cluster ``C'' lets the material discriminate between $\gamma_\text{read} = 3\%$ and $\gamma_\text{read} = 4\%$. When the strain amplitude is reduced from 4\% to 3\%, cluster ``C''   stops switching states, and doesn't resume until $\gamma_\text{read} \ge 3.5\%$. Similarly, cluster ``A'' distinguishes $\gamma_\text{read} \le 4\%$, contributing to the readout of the 4\% memory. Cluster ``D''  distinguishes among values of $\gamma_\text{read}$, but it ends every cycle in the same state --- it is unused by our readout method. Finally, cluster ``E'' is nearly latent until $\gamma_\text{read} > 4\%$, and so reports the largest amplitude during training. 

% In separate experiments, we have confirmed that there is a population of rearrangements that is activated when we exceed the largest training amplitude, and that do not revert even when smaller-amplitude shearing is resumed. It is possible that these rearrangements are truly irreversible (i.e.\ cannot be reverted by \emph{any} bulk deformation), analogous to the irreversible changes in more dilute suspensions when the largest training amplitude is exceeded~\cite{Keim:2011dv,Paulsen:2014hm}.

\subsection{Discussion}

By observing the motions of particles, and considering a simple example of RPM, we have shown how our material's memory arises from the hysteresis of individual rearranging clusters, each of which responds differently to global deformations. Hysteresis is responsible for the non-monotonic readout curves in Figs.~\ref{fig:readexp}c and \ref{fig:read2}b, and explains why this behavior is different from dilute suspensions (Fig.~\ref{fig:readexp}d), in which the steady state exhibits kinematic reversibility, not hysteresis.
Our results raise the question of how this behavior is connected with the physics of amorphous solids. A single rearranging cluster has hysteresis and is coupled to external shear stress, analogous to a hysteron in the Preisach model. However, it is also coupled to elastic deformations of the surrounding material~\cite{Eshelby:1957dg,Keim:2014hu}, so it may interact with nearby clusters, violating an assumption of the Preisach model. Indeed, when we measure the $\gamma^+_i$ and $\gamma^-_i$ (analogous to $H^+_i, H^-_i$), we find they depend on strain amplitude, presumably due to other, nearby rearrangements becoming active or inactive as the amplitude is varied (Fig.~\ref{fig:activity}b). RPM is proven to hold exactly only when interactions are ``ferromagnetic'' (each rearrangement encourages others) \cite{Sethna:1993ts}, but here we can also have ``antiferromagnetic'' interactions, depending on the relative positions of rearranging clusters~\cite{Eshelby:1957dg,Mungan2019}. Instead of the Preisach model, we can look to studies of disordered magnetic systems more generally, where despite complex, frustrated interactions, RPM may still hold at least approximately~\cite{deutsch04, Pierce:2005dd, Hovorka:2008ks}, especially in a steady state under cyclic driving~\cite{Hovorka:2008ks, Gilbert:2015db,  Mungan:2019fh,Mungan2019}. 

In the magnetic systems just discussed, disorder is quenched---the Hamiltonian prescribes couplings of a fixed population of subsystems to each other and an external field---facilitating the return to previous states. But disorder in deformed solids is generally \emph{not} quenched~\cite{Priezjev:2013hp, Keim:2014hu, Nagamanasa:2014jx, Regev:2015hs}. Instead, the transient at the beginning of each experiment remodels the material irreversibly, until we are left with a stable population of repeating rearrangements~\cite{Regev:2013es,Keim:2014hu,Regev:2015hs}. Remarkably, even as we subsequently reduce the strain amplitude and change the state of the system, this population largely persists~\cite{Mungan2019}. The few outlier trials we discard from our analysis~\cite{suppmat} may be exceptions.
%The transient is outside the scope of this paper, but we can see that by shaping that population it may encode information in one more way: comparing the ``4'' curve in Fig.~\ref{fig:read2}b with the ``3,~4'' curve suggests that these differently-prepared systems are subtly different, while RPM with permanently quenched disorder would make them indistinguishable.

% Disorder in our experiments is not quenched but it is nearly \emph{quiescent}. (Perfect quiescence is not needed: Fiocco \etal~\cite{Fiocco:2014bz} found that multiple memories could be read late in the transient, and in the present experiments, external vibrations and slow aging of the interface cause small intermittent changes.) 

While it exactly describes the behavior of only a few kinds of systems~\cite{Keim:2019aa}, here return-point memory is a generic prototype of how a rich global memory behavior can arise from disorder and local hysteresis (i.e. metastability). This suggests that the kind of memory discussed here might not only be present in the many kinds of amorphous solids, but could also be found or even engineered in many other types of systems~\cite{Keim:2019aa, Laurson:2012bc,Slotterback:2012fa, Ren:2013cp, Royer:2015dj,Dobroka:2017fc} given appropriate driving.  Finally, by illuminating the mechanism for this behavior, our work points to a more precise question: why our material's self-organized steady states, despite frustration and marginal stability~\cite{Regev:2015hs,Mungan:2019aa,Mungan2019}, are so amenable as we vary driving to retrieve memories.

% The steady state of amorphous solids under cyclic driving has been variously called a ``limit cycle,'' ``periodic,'' ``cyclic,'' ``reversibly plastic,'' and ``loop-reversible'' by us and other authors. Each term describes the motions of the particles under periodic shear with constant amplitude, implying that once that amplitude is reduced and the periodicity is broken, the steady state ends. But we have identified an aspect of the system that persists: the hysteretic subsystems that make it possible to recall that same ``limit cycle'' at a later time, and even nest one limit cycle inside another, as in Fig.~\ref{fig:activity}b. Return-point memory is the prototype for how disorder and hysteresis create these memories, and it . Our results challenge the prevalent conception of these steady states, and may shed light on the physics of marginal stability that underlie them~\cite{Regev:2015hs}.

\begin{acknowledgments}
For illuminating discussions we thank Muhittin Mungan, Karin Dahmen, Sidney Nagel, Srikanth Sastry, Ajay Sood, and Joseph Paulsen. We also thank Paulo Arratia, David Gagnon, Larry Galloway, Luke Horowitz, Dani Medina, and Peter Nelson for their help. NCK thanks the Kavli Institute for Theoretical Physics for its hospitality, supported in part by NSF grant PHY-1748958. Minus K Technology donated the 100BM-1 vibration isolation platform used in experiments. This work was supported by NSF grant DMR-1708870, by an RSCA grant from California Polytechnic State University, San Luis Obispo, and by the William and Linda Frost Fund.
\end{acknowledgments}

\bibliography{references,references-misc}

\end{document}